# Charge Transfer Dynamics in MoSe$_2$/hBN/WSe$_2$ Heterostructures

Yoseob Yoon[1,2,*], Zuocheng Zhang[1], Ruishi Qi[1,2], Andrew Y. Joe[1,2], Renee Sailus[3],
Kenji Watanabe[4], Takashi Taniguchi[5], Sefaattin Tongay[3], and Feng Wang[1,2]

[1]Department of Physics, University of California, Berkeley, California 94720, United
States
[2]Materials Sciences Division, Lawrence Berkeley National Laboratory, Berkeley, California
94720, United States
[3]School for Engineering of Matter, Transport and Energy, Arizona State University,
Tempe, Arizona 85287, United States
[4]Research Center for Functional Materials, National Institute for Materials Science, 1-1
Namiki, Tsukuba 305-0044, Japan
[5]International Center for Materials Nanoarchitectonics, National Institute for Materials
Science, 1-1 Namiki, Tsukuba 305-0044, Japan

[*]Correspondence to Yoseob Yoon (yoonys@berkeley.edu)

# Abstract

Ultrafast charge transfer processes provide a facile way to create interlayer excitons in directly contacted transition metal dichalcogenide (TMD) layers. More sophisticated heterostructures composed of TMD/hBN/TMD enable new ways to control interlayer exciton properties and achieve novel exciton phenomena, such as exciton insulators and condensates, where longer lifetimes are desired. In this work, we experimentally study the charge transfer dynamics in a heterostructure composed of a 1 nm thick hBN spacer between $MoSe_2$ and $WSe_2$ monolayers. We observe the hole transfer from $MoSe_2$ to $WSe_2$ through the hBN barrier with a time constant of 500 ps, which is over 3 orders of magnitude slower than that between TMD layers without a spacer. Furthermore, we observe strong competition between the interlayer charge transfer and intralayer exciton-exciton annihilation processes at high excitation densities. Our work opens possibilities to understand charge transfer pathways in TMD/hBN/TMD heterostructures for the efficient generation and control of interlayer excitons.

# Abstract Graphic

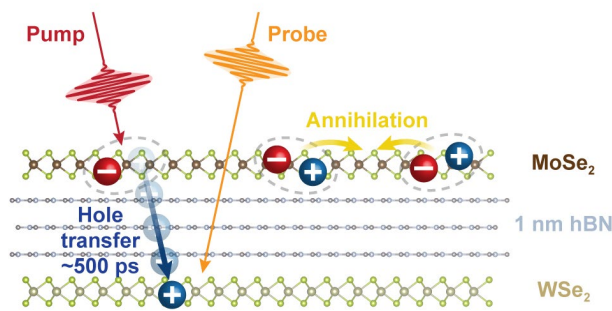

Interlayer excitons are bound electron-hole pairs where the constituent electron and hole are spatially separated near the junction of two layered semiconductors. They are energetically favorable in a heterobilayer with type-II band alignment or in a homobilayer with an out-of-plane electric field. In directly contacted transition metal dichalcogenide (TMD) heterostructures, interlayer excitons can be easily generated by any optical excitations above one of the intralayer exciton energies, owing to ultrafast and efficient charge transfer processes from one layer to the other.[1-4]

Adding a thin hexagonal boron nitride (hBN) layer to make a TMD/hBN/TMD heterostructure provides the opportunity to engineer many properties of the interlayer excitons, such as their binding energy, dipolar exciton-exciton interaction strength, and charge transfer dynamics. Since electrons reside in one layer and holes reside in the other layer that are separated by the hBN layer, the spatial overlap between their wave functions is poor and their radiative lifetimes can be much longer[5] than those of directly contacted TMD/TMD heterostructures.[6-9] A long lifetime is one of the key properties that allows us to reach novel excitonic phases in equilibrium, including excitonic insulators[5,10,11] and Bose-Einstein condensates,[12] where excitons need to interact for a sufficiently long time before they recombine and disappear. In addition, the aligned nature of the permanent dipole moments of interlayer excitons leads to enhanced exciton-exciton interaction strength with increased hBN thickness.[13] The dipolar interaction strength of interlayer excitons can exceed the exchange interaction between neutral excitons, which can be used to stabilize strongly correlated excitonic phases.[5,14,15]

Although the hBN layer is beneficial for realizing long-lived and strongly correlated interlayer excitons, it exponentially suppresses the tunneling probability and the charge transfer rate. This will hamper the generation of such interlayer excitons through optical excitations. An experimental study of these effects in TMD/hBN/TMD heterostructures would be crucial for understanding the dynamics of interlayer exciton formation and for optimizing the strongly correlated excitonic phases.

Transient absorption spectroscopy is a powerful tool to investigate the interlayer exciton dynamics, where pump-induced changes of absorption peaks are probed as a function of the time delay between pump and probe pulses. Since the intralayer exciton peak changes (induced by the presence of interlayer excitons) are probed in these measurements, the vanishing oscillator strength of interlayer excitons[7] does not prohibit the measurement; for this reason, transient absorption spectroscopy has been primarily used to study the dynamics of charge transfer processes.[1,16–22] Here, using the same technique, we investigate the ultrafast charge transfer dynamics in a heterostructure composed of a 1 nm thick hBN spacer between $MoSe_2$ and $WSe_2$ monolayers. We reveal that the hole transfer from $MoSe_2$ to $WSe_2$ across hBN takes place with a time constant of ∼500 ps, which is over 3 orders of magnitude slower than the charge transfer between directly contacted TMD layers (<100 fs). In addition, we show that there is a strong competition between the interlayer charge transfer and intralayer exciton-exciton annihilation processes in this system at high excitation densities.

Figure 1a,b illustrates our experimental scheme. We first create $MoSe_2$ intralayer excitons using a quasiresonant pump pulse, which essentially does not create $WSe_2$ excitons ($E_{pump} \approx E_{MoSe_2} < E_{WSe_2}$). After a time delay ($\Delta t = t_{probe} - t_{pump}$), a broadband probe pulse (1.55-1.91 eV) measures the $WSe_2$ intralayer exciton absorption peak. Holes dissociated from the $MoSe_2$ excitons are transferred to the $WSe_2$ layer and relaxed to its band edge, which saturate the $WSe_2$ intralayer exciton absorption by Pauli blocking. A microscope image and schematic layered structure are shown in Figure 1c,d, respectively. The hBN spacer consists of three atomic layers, which is confirmed by an atomic force microscope (AFM) measurement (see Figure S1a,b). The twist angle between $MoSe_2$ and $WSe_2$ layers is determined to be 19.5° by a polarization-dependent second-harmonic generation measurement (see Figure S1c).

Due to limitations in typical experimental configurations (e.g., opaque substrate or cryostat sample holder, see Figure S2), it is common that only the reflection is measured. To separate the optical features of the sample ($MoSe_2$/hBN/$WSe_2$) from the background ($SiO_2$/Si substrate and other dielectric layers) and to eliminate the effect of the uneven probe spectrum, we measure a reflection contrast spectrum, defined as $(I_r^{sb} - I_r^b)/I_r^b$, where $I_r^{sb}$ is a reflection spectrum measured in a region with sample and background ("sb"), and $I_r^b$ is a reflection spectrum measured in a background ("b") region, as defined in Figure 1d. The measured reflection contrast and pump spectra are shown in Figure 1e. The pump is mostly resonant with MoSe$_2$ excitons, although it is slightly red-shifted to reduce the direct generation of WSe$_2$ excitons and carriers. The small mismatch is taken into account in the calibration of pump absorption by the MoSe$_2$ layer (see Figure S4).

The local field at the TMD layers can be vastly different from that of a free-space field due to the interference between reflections from multiple interfaces in the layered structure. Depending on the structure's dimensions and dielectric properties, the Fano-like line shapes and peak heights of resonances can dramatically vary even with the same TMD layer. Therefore, it is crucial to perform transfer-matrix calculations to estimate the absolute values of absorption (see the Supporting Information for details of the calibration). A result of the transfer-matrix fitting to the measured reflection contrast spectrum, starting from known thicknesses and dielectric constants of each layer, is shown as the orange line in Figure 1e. From the absolute values of the reflection contrast spectrum, not only are the central energies and line widths extracted but also the absolute values of absorption by each TMD layer are determined (see Figure S4), which cannot be simply read from the reflection contrast values.

Figure 2a shows representative transient reflection spectra measured by the probe with and without the pump, which we denote as $R_{\text{pump on}}$ and $R_{\text{pump off}}$, respectively. The pump-on spectrum is taken with a high pump fluence ($F_{\text{pump}} = 39.4$ μJ/cm$^2$) and at a pump-probe time delay when the hole transfer is completed ($\Delta t = 30$ ps). A probe spectrum in the background region is also measured to obtain the reflection contrast spectra shown in Figure 2b. At this high pump fluence, the exciton absorption peak saturates and the trion absorption peak ($\sim$20 meV below the exciton energy) emerges. Transfer-matrix calculations including these two peaks can successfully reproduce the measured reflection contrast spectra shown in Figure 2b, and the extracted absorption profiles are shown in Figure 2c.

We then varied the time delay to measure the charge transfer dynamics. Figure 3a,b shows the transient reflection spectra at low and high pump fluences, respectively. To visualize small changes in the reflection intensities, we plotted their differential changes, namely, $\Delta R/R = (R_{\text{pump on}} - R_{\text{pump off}})/R_{\text{pump off}}$. In stark contrast to pump-probe spectra measured on similar devices without any hBN spacer,[1] Figure 3a shows that the WSe$_2$ exciton peak at 1.6923 eV rises slowly and saturates by $\sim$100 ps when the pump fluence is low, which corresponds to the initial MoSe$_2$ exciton density $n_{\mathrm{m}}(\Delta t = 0) = 4.6 \times 10^{11}$ cm$^{-2}$. When the initial MoSe$_2$ exciton density is increased to $n_{\mathrm{m}}(0) = 2.7 \times 10^{12}$ cm$^{-2}$, the WSe$_2$ exciton peak saturates by $\sim$30 ps, as shown in Figure 3b. The observed density-dependent dynamics suggests that nonlinear interactions are involved in the charge transfer process in the MoSe$_2$/hBN/WSe$_2$ heterostructure. The vertical line cuts of transient reflection spectra at various time delays are shown in Figure 3c,d, which reveal both the saturation of the exciton peak ($\Delta R/R > 0$) and the appearance of a trion peak ($\Delta R/R < 0$).

To quantitatively measure physical properties, we analyzed the reflection contrast spectrum at each time delay and fitted to the transfer-matrix calculation result. We used the Lorentzian oscillator model with complex dielectric constants to approximate the WSe$_2$ absorption resonance and used the oscillator strength $f$, central energy $E$, and line width $\Gamma$ as fitting parameters at each time delay. Other parameters are fixed to the values obtained from the fitting without the pump (the fit result shown in Figure 1e). The time traces of the extracted quantities are plotted in Figure 4. We calculated $\Delta A/A = (A_{\mathrm{pump\ on}} - A_{\mathrm{pump\ off}})/A_{\mathrm{pump\ off}}$, where $A$ is the absolute value of the absorption, as well as the pump-induced line width broadening $\Delta \Gamma = \Gamma_{\mathrm{pump\ on}} - \Gamma_{\mathrm{pump\ off}}$ and pump-induced energy shift $\Delta E = E_{\mathrm{pump\ on}} - E_{\mathrm{pump\ off}}$. Excitation-density-dependent saturation dynamics are observed (Figure 4a), and the line width broadening dynamics follows a similar trend (Figure 4b). The absorption saturation and line width broadening effects can be understood as a result of the Pauli blocking at the WSe$_2$ exciton state by the transferred holes, and these effects are consistent with the repulsive polaron picture.[23] On the other hand, the peak energy slightly red-shifts immediately at $\Delta t = 0$ and then blue-shifts at later time delays. We attribute the initial red-shift to the dielectric screening effect from carriers in the MoSe$_2$ layer; thus, the effect occurs instantaneously after the optical excitation and does not reflect the charge transfer dynamics. After the carrier-induced screening decays, the transferred holes result in the blue-shift of the WSe$_2$ exciton peak (as well as its saturation and broadening). We emphasize that such spectral analysis and transfer-matrix calculations are crucial to deconvolute various processes from complicated transient reflection spectra and to obtain meaningful results.

From a separate gate-dependent reflection contrast measurement (see Figure S5), we were able to correlate the density of transferred holes in the WSe$_2$ layer, $n_\mathrm{w}$, and the change of WSe$_2$ exciton absorption induced by hole doping. This way, we can convert the $\Delta A/A$ data (as in Figure 4a) to $n_\mathrm{w}$, which is plotted in Figure 5a as a function of the time delay and excitation density. The estimated density of MoSe$_2$ excitons created by the pump, $n_\mathrm{m}(\Delta t = 0)$, ranges from $5.6 \times 10^{10}$ to $5.5 \times 10^{12}$ cm$^{-2}$ (see the Supporting Information for details of the calibration). The hole transfer time scale is not only much slower than that without an hBN spacer but also strongly density dependent. Due to the slower time scale, it competes with several other processes at the picosecond time scale. Most importantly, the exciton-exciton annihilation process within the MoSe$_2$ layer becomes very important at higher pump fluences.[24,25]

The maximum density of transferred holes, $\max[n_\mathrm{w}(\Delta t)]$, is plotted in Figure 5b, and the charge transfer efficiency, defined as $\max[n_\mathrm{w}(\Delta t)]/n_\mathrm{m}(0)$, is plotted in Figure 5c. As the excitation density is increased, there is increased competition between the charge transfer and other nonlinear processes, which leads to a less efficient transfer. This behavior is different from devices without an hBN spacer,[16] where the charge transfer process dominates all other losses and a linear increase of $\max[n_\mathrm{w}(\Delta t)]$ is observed as a function of $n_\mathrm{m}(0)$ until the transferred hole density becomes too large. In a similar device but with a 3 nm thick hBN spacer, the charge transfer becomes too slow and inefficient, so we were not able to measure any appreciable change of the WSe$_2$ exciton peak within the measurement window (500 ps).

To quantitatively understand the density-dependent dynamics and competition between different processes, we used coupled rate equations to solve for the exciton and hole densities. The exciton density in the MoSe$_2$ layer is denoted by $n_\mathrm{m}$, while the transferred hole density in the WSe$_2$ layer is denoted by $n_\mathrm{w}$. The rates of their population change can be expressed as

$$\frac{dn_\mathrm{m}}{dt} = -\gamma_\mathrm{m} n_\mathrm{m} - \frac{1}{2}\gamma_\mathrm{mm} n_\mathrm{m}^2 - \gamma_\mathrm{t} n_\mathrm{m} \qquad (1)$$

$$\frac{dn_\mathrm{w}}{dt} = -\gamma_\mathrm{w} n_\mathrm{w} + \gamma_\mathrm{t} n_\mathrm{m} + \frac{1}{2}\alpha\gamma_\mathrm{mm} n_\mathrm{m}^2 \qquad (2)$$

where $\gamma_\mathrm{m}$ is the MoSe$_2$ exciton population decay rate, $\gamma_\mathrm{w}$ is the WSe$_2$ hole population decay rate, $\gamma_\mathrm{mm}$ is the MoSe$_2$ exciton-exciton annihilation rate, and $\gamma_\mathrm{t}$ is the hole transfer rate from MoSe$_2$ to WSe$_2$. The last term in Eq. (2) represents a nonlinear charge transfer mechanism, where the high-energy excitations created by the exciton-exciton annihilation provide additional transfer pathways.

We numerically solved the coupled differential equations with fitting parameters to reproduce the measured density-dependent dynamics and plotted it in Figure 5a as gray dotted lines. The global parameters that give best fits to all excitation densities are $\gamma_m = (100 \text{ ps})^{-1}$, $\gamma_w = (4 \text{ ns})^{-1}$, $\gamma_{mm} = 0.3 \text{ cm}^2/\text{s}$, and $\gamma_t = (500 \text{ ps})^{-1}$ (see Figure S8 for fit results using different starting parameters). The value of $\gamma_{mm}$ is consistent with that observed in a previous study with a MoSe$_2$ monolayer,[25] and the intralayer exciton decay rate $\gamma_m$ represents the effective rate that includes radiative decay, nonradiative decay, and the dynamics between bright and dark excitonic species[26] (see Figure S7 for a separate measurement). Without the nonlinear charge transfer term $\frac{1}{2}\alpha\gamma_{mm}n_m^2$, the very steep saturation at higher excitation densities cannot be reproduced with any reasonable fit parameter values. We estimate that $\alpha = 0.023$, which implies that about 2.3% of the annihilated particles eventually find a way to the WSe$_2$ layer.

The hole transfer time of 500 ps is over 3 orders of magnitude longer than that observed in a directly contacted MoSe$_2$/WSe$_2$ heterostructure ($<100$ fs). The phonon relaxation time to the band edge is negligibly short (optical phonons at ~40 meV can be emitted efficiently at the femtosecond time scale[27–29]), so the charge transfer process is only slowed down by the increased tunneling barrier and reduced interlayer coupling. However, even though the transfer time scale is much slower, the overall transfer efficiency is still significant, so the optical generation of interlayer excitons is minimally compromised.

The measured efficiency shown in Figure 5c gives additional bounds to the fitting result. In the low excitation-density limit, the analytical solutions of Eqs. (1) and (2) show that the transfer efficiency should be $\eta = \gamma_t/(\gamma_t + \gamma_m)$. The measured efficiency of 16% at the lowest excitation density implies that the transfer time $\tau_t$ is about 5 times longer than the exciton lifetime $\tau_m$, which is consistent with the measurements and fit results. Note that, without the density-dependent measurements, the lowest-density data alone cannot satisfactorily distinguish the charge transfer time $\tau_t$ from other time scales on the same order. In addition, we exclude the possibility of faster charge transfer at higher excitation densities, which would result in higher transfer efficiencies at higher excitation densities. Therefore, we confirm that the charge transfer rate is indeed hundreds of picoseconds in $MoSe_2$/1 nm hBN/$WSe_2$.

In conclusion, we used a thin hBN spacer between $MoSe_2$ and $WSe_2$ monolayers to study the effects of the tunneling barrier on the charge transfer time scale and efficiency. We extracted pump-induced saturation, energy shift, and line width broadening at each time delay to distinguish different processes that affect the absorption peak. Density-dependent saturation dynamics reveals competing processes such as exciton-exciton annihilation and nonlinear charge transfer processes. By fitting the solution of coupled rate equations to the measured density-dependent charge transfer dynamics, we found that the intrinsic charge transfer time is ~500 ps, which is over 3 orders of magnitude slower than without a spacer. Our work opens possibilities of investigating the factors contributing to charge transfer, e.g., the effects of momentum mismatch between the layers, which have not been feasible in directly contacted TMD heterostructures due to the extremely fast time scale. Understanding the charge transfer process across various thicknesses of an hBN spacer will allow us to further optimize the generation and control of interlayer excitons with tunable dipole moments and lifetimes.

## Supporting Information

The Supporting Information is available free of charge at https://pubs.acs.org.

Device fabrication methods and characterization results, pump-probe setup and beam profiles for the pump fluence determination, detailed calibration methods for the $MoSe_2$ exciton density using transfer-matrix calculation of absorption, detailed calibration methods for the $WSe_2$ hole density using gate-dependent absorption measurement, results of coupled rate equations fitting with different conditions

## Acknowledgments

This work was primarily supported by the Center for Computational Study of Excited-State Phenomena in Energy Materials (C2SEPEM) at LBNL, funded by the U.S. Department of Energy, Office of Science, Basic Energy Sciences, Materials Sciences and Engineering

Division under Contract No. DE-AC02-05CH11231. The device fabrication is supported by the U.S. Department of Energy, Office of Science, Basic Energy Sciences, Materials Sciences and Engineering Division under Contract No. DE-AC02-05CH11231 within the van der Waals heterostructure program (KCWF16). R.S. and S.T. acknowledge primary support from DOE-SC0020653 (materials synthesis), NSF CMMI 1825594 (NMR and TEM studies), NSF DMR-1955889 (purity measurements), NSF CMMI-1933214, NSF 1904716, NSF 1935994, NSF ECCS 2052527, DMR 2111812, and CMMI 2129412 (for defect-structure-quality relations). K.W. and T.T. acknowledge support from the JSPS KAKENHI (Grant Numbers 19H05790 and 20H00354). We thank Y. R. Shen for helpful discussion.

## Author contributions

F.W. and Y.Y. conceived the project. Y.Y. performed pump-probe measurements and transfer-matrix calculations. Z.Z., R.Q., and A.Y.J. fabricated the samples and performed gate- and temperature-dependent reflection contrast measurements. K.W. and T.T. grew hBN crystals. R.S. and S.T. grew $MoSe_2$ and $WSe_2$ crystals. All the authors discussed the results and contributed to the manuscript.

## Notes

The authors declare no competing financial interest.

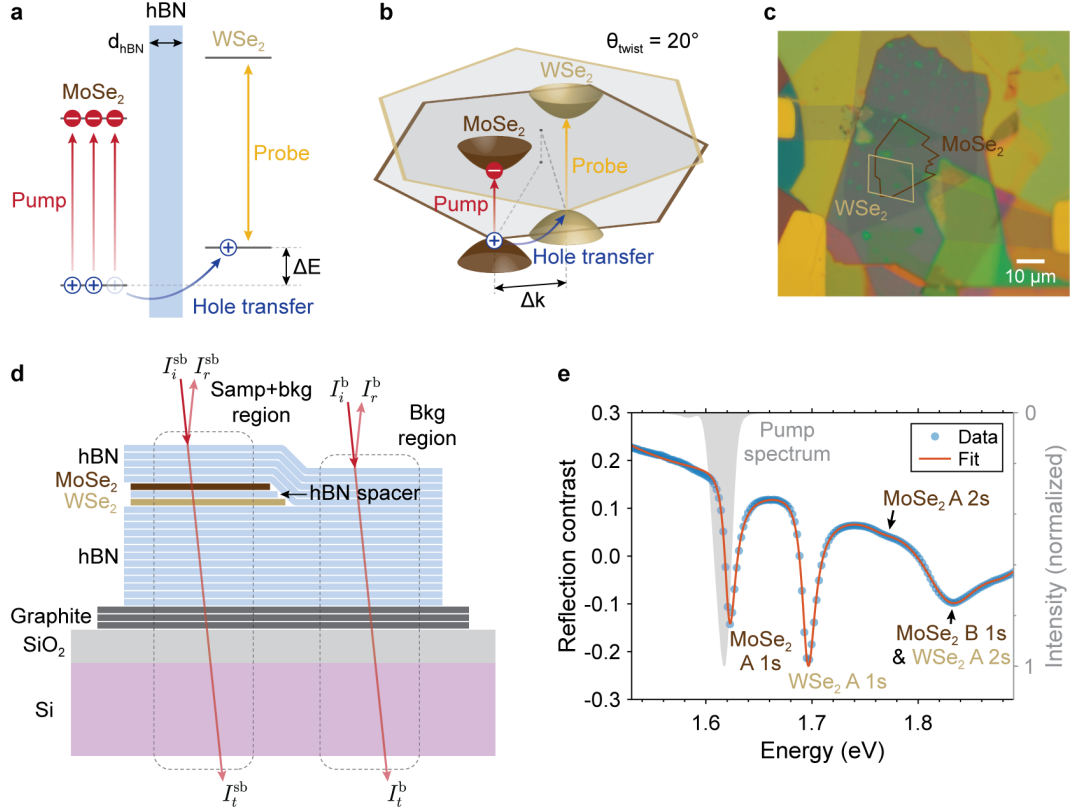

**Figure 1.** Charge transfer pathway and pump-probe measurement scheme. (a) Charge transfer pathway in real space. A quasiresonant pump pulse creates MoSe₂ excitons (red arrows). Holes in the MoSe₂ layer are transferred to the WSe₂ layer (blue arrow). The WSe₂ exciton absorption spectrum is monitored by a broadband probe pulse (yellow arrow). The transfer process is accompanied by a real-space displacement of holes by 1.6 nm ($d_{\mathrm{hBN}} = 1.0$ nm) and energy relaxation by $\Delta E \approx 0.3$ eV. (b) Charge transfer pathway in momentum space. For the device with $\theta_{\mathrm{twist}} = 20°$, the transfer process is accompanied by a momentum shift of $\Delta k \approx 0.44$ Å$^{-1}$. (c) A microscope image of the MoSe₂/hBN/WSe₂ heterostructure. Graphite gates and contacts (gray shaded area) are grounded unless otherwise noted. (d) Schematic of the layered structure. Incident pulse intensity $I_i$ is reflected ($I_r$), transmitted ($I_t$), and absorbed by the TMD layers, while only the reflected part is measured in the sample + background ("sb") region and the background ("b") region. (e) Reflection contrast spectrum at 100 K (blue circles) and a fit (orange line) using the transfer-matrix simulation described in the main text. A normalized pump spectrum is shown as the gray shaded area (inverted vertical axis on the right-hand side).

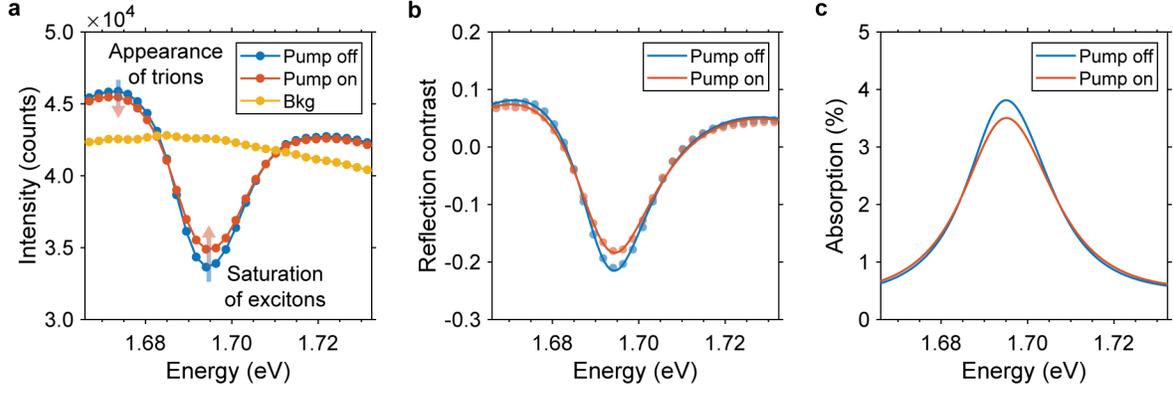

**Figure 2.** Determination of WSe$_2$ absorption spectra with and without the pump at 100 K. (a) Raw reflection intensities of the broadband probe in the background region (yellow), in the sample region without the pump (blue), and in the sample region with the pump (orange). The pump fluence was $F_{\text{pump}} = 39.4$ µJ/cm$^2$, corresponding to the initial MoSe$_2$ exciton density $n_{\text{m}}(\Delta t = 0) = 3.0 \times 10^{12}$ cm$^{-2}$. The pump-on spectrum is taken at a time delay of 30 ps, when the hole transfer process is completed. (b) Reflection contrast spectra with and without the pump (blue and orange circles, respectively). The best fits to the transfer-matrix calculations are shown as blue and orange lines. (c) Extracted absorption profiles with and without the pump (blue and orange lines, respectively).

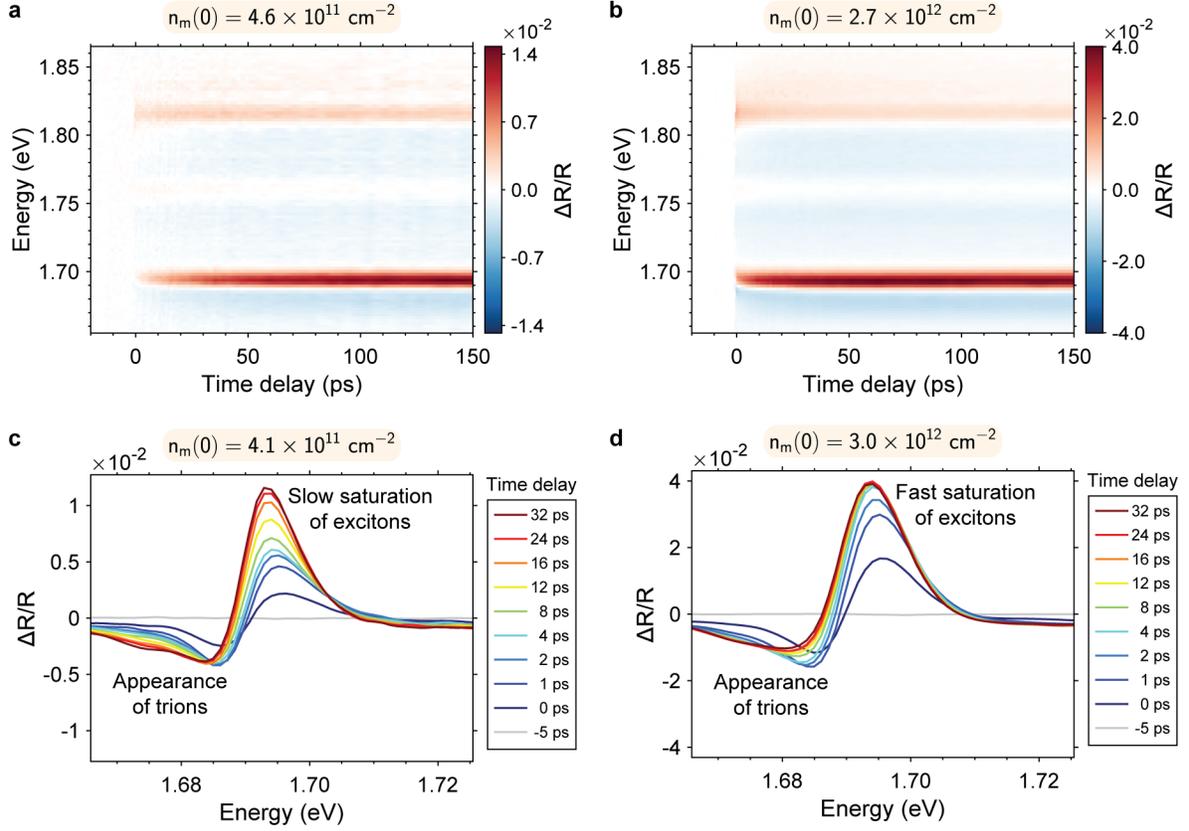

**Figure 3.** Pump-probe spectra at 100 K and transient spectral analysis. (a,b) Differential transient reflection spectra ($\Delta R/R$) as a function of the probe energy and time delay at low and high pump fluences. (c,d) Line cuts of transient reflection spectra at various time delays (different colors) and at low and high pump fluences.

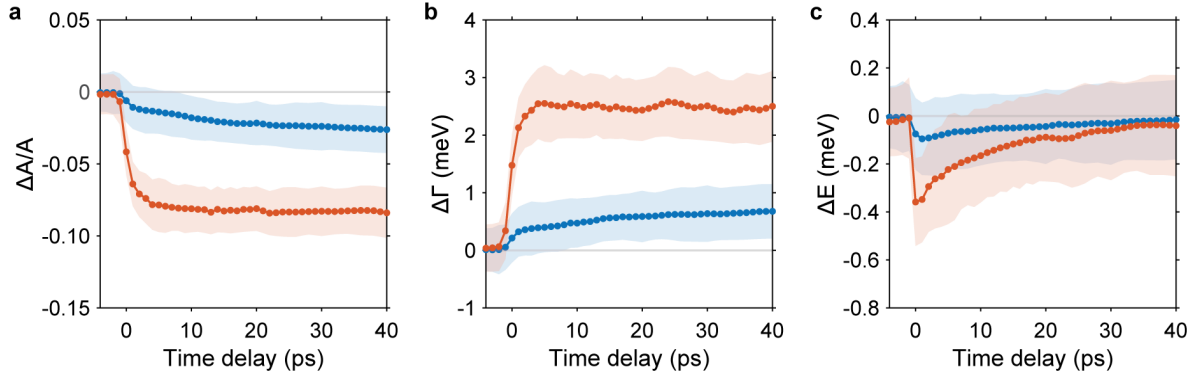

**Figure 4.** Transient spectral analysis as a function of time delay. Absolute values of (a) pump-induced absorption change $\Delta A/A$, (b) pump-induced line width broadening $\Delta\Gamma$, and (c) pump-induced energy shift $\Delta E$, when the initial MoSe$_2$ exciton densities are $n_\mathrm{m}(0) = 4.1 \times 10^{11}$ cm$^{-2}$ (blue) and $n_\mathrm{m}(0) = 3.0 \times 10^{12}$ cm$^{-2}$ (orange). Filled circles are the best-fit parameters from the transfer-matrix calculation at each time delay, and shaded regions denote the 90% confidence interval from the fitting.

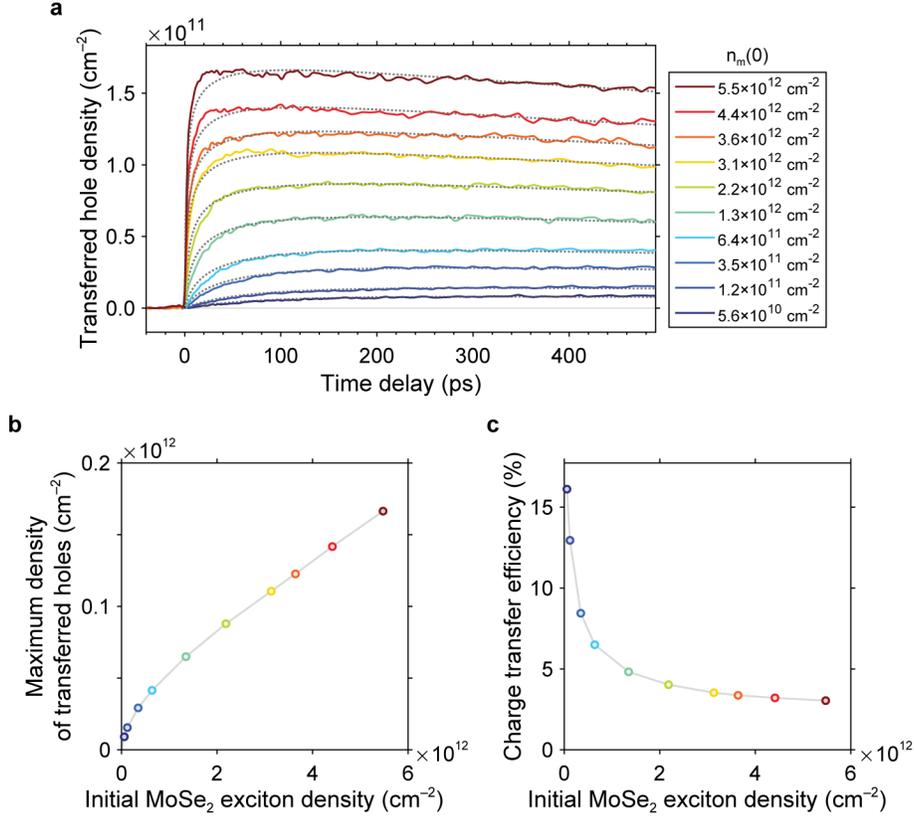

**Figure 5.** Charge transfer dynamics as a function of the initial MoSe₂ exciton density at 100 K. (a) Transferred hole density in the WSe₂ layer is plotted as a function of time delay for different values of the initial MoSe₂ exciton density, $n_{\mathrm{m}}(0)$. Black dotted lines are fit results using Eqs. (1) and (2). (b) The maximum transferred hole density, $\max[n_{\mathrm{w}}(\Delta t)]$, as a function of the initial MoSe₂ exciton density, $n_{\mathrm{m}}(0)$. (c) The charge transfer efficiency, defined as $\max[n_{\mathrm{w}}(\Delta t)]/n_{\mathrm{m}}(0)$, as a function of $n_{\mathrm{m}}(0)$.

# Supporting Information for
# "Charge Transfer Dynamics in MoSe$_2$/hBN/WSe$_2$ Heterostructures"

Yoseob Yoon,[1, 2, *] Zuocheng Zhang,[1] Ruishi Qi,[1, 2] Andrew Y. Joe,[1, 2] Renee Sailus,[3]

Kenji Watanabe,[4] Takashi Taniguchi,[5] Sefaattin Tongay,[3] and Feng Wang[1, 2]

[1] *Department of Physics, University of California, Berkeley, California 94720, United States*

[2] *Materials Sciences Division, Lawrence Berkeley National Laboratory, Berkeley, California 94720, United States*

[3] *School for Engineering of Matter, Transport and Energy,*
*Arizona State University, Tempe, Arizona 85287, United States*

[4] *Research Center for Functional Materials, National Institute for Materials Science, 1-1 Namiki, Tsukuba 305-0044, Japan*

[5] *International Center for Materials Nanoarchitectonics,*
*National Institute for Materials Science, 1-1 Namiki, Tsukuba 305-0044, Japan*

## Contents

## 1. Device fabrication and characterization

The hBN-encapsulated MoSe$_2$/hBN/WSe$_2$ heterostructure was fabricated using a dry transfer technique based on polypropylene carbonate (PPC). The exfoliated MoSe$_2$ and WSe$_2$ monolayers are separated by a thin hBN layer with a thickness of around 1 nm. Two additional few-layer graphenes (FLG) are in contact with the MoSe$_2$ and WSe$_2$ monolayers, respectively, and the structure is encapsulated by the bottom and top hBN layers. The bottom and top gates are also made of FLG. The stack is finally released on a 285 nm SiO$_2$/Si substrate. A standard photolithography system (Durham Magneto Optics, MicroWriter) is used to define the electrodes, and an $e$-beam deposition system is used for the 5 nm Cr/100 nm Au deposition. The WSe$_2$ monolayer, MoSe$_2$ monolayer, top and bottom gates, and heavily hole-doped Si are grounded during the measurements unless otherwise noted.

The spacer hBN thickness is crucial for controlling the interlayer coupling and charge transfer dynamics between MoSe$_2$ and WSe$_2$ layers. We characterized the thickness using an atomic force microscope (AFM), as shown in Figure S1. A linecut across the boundary, denoted as the black line in Figure S1a, is fit to a function

$$z(x) = \frac{d_{\mathrm{hBN}}}{2} \operatorname{erfc}\left(\frac{x - x_0}{w}\right) + ax^2 + bx + c \tag{S1}$$

where $z$ is the height, $x$ is the position, $x_0$ is the position of the hBN edge, $d_{\mathrm{hBN}}$ is the hBN thickness, $\operatorname{erfc}(x)$ is the complementary error function, $w$ is the width of the height increment at the boundary due to the finite tip radius, and $ax^2 + bx + c$ is a smooth quadratic background function. Figure S1b shows the background-subtracted result. The fitted hBN thickness is 0.92 nm, which is consistent with the thickness of three hBN layers (the layer-to-layer distance of the hBN crystal is 0.333 nm).

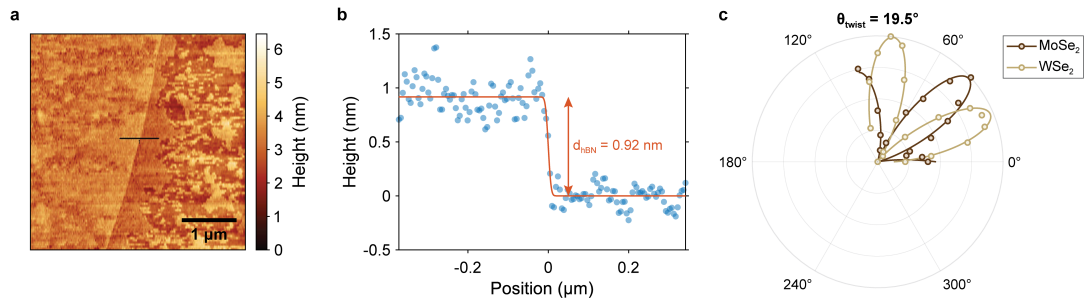

**Figure S1.** Characterization of the hBN thickness. (a) AFM topography of the spacer hBN. (b) A background-subtracted linecut across the boundary (blue circles) and a fit result (orange line). (c) Polarization-dependent second-harmonic intensities in the monolayer regions.

---

\* yoonys@berkeley.edu

The twist angle between MoSe$_2$ and WSe$_2$ layers is determined by the second-harmonic generation (SHG) method. We used femtosecond laser pulses at $\lambda = 800$ nm and collected the generated second-harmonic intensity at $\lambda = 400$ nm, while rotating the incident linear polarization using a half waveplate. Measured polarization-dependent second-harmonic intensities in MoSe$_2$ and WSe$_2$ monolayer regions are shown in Figure S1c. From the fit results, the twist angle between MoSe$_2$ and WSe$_2$ layers in the MoSe$_2$/1 nm hBN/WSe$_2$ device used in the main text is determined to be 19.5°. The second-harmonic intensity in the heterostructure region was greater than that in the monolayer regions, which indicates that the twist angle is 19.5° rather than 40.5°.

## 2. Pump-probe setup and beam profiles

Yb-based femtosecond laser (Light Conversion CARBIDE, wavelength $\lambda = 1030$ nm, repetition rate $f_{\mathrm{rep}} = 151.3$ kHz) was used as a "fundamental" beam for an optical parametric amplifier (OPA, Light Conversion ORPHEUS). The OPA signal energy was tuned to be resonant with the MoSe$_2$ exciton energy ($E_{\mathrm{pump}} = 1.6165$ eV, $\lambda_{\mathrm{pump}} = 767$ nm) and used as a "pump" beam. The leftover fundamental beam was focused onto a 3-mm-thick sapphire crystal to generate a supercontinuum, which was filtered to let through $E_{probe} = 1.55\text{-}1.91$ eV and used as a "probe" beam. The pulse durations (full width at half maximum, FWHM, of the intensity envelope) of both pump and probe pulses were initially 200 fs but elongated to 250 fs after passing through multiple optical components, including an objective lens (20×, NA 0.45). The FWHM of the pump beam's spatial profile was 6.2 μm, while that of the probe beam was 1.6 μm (see Figure S3). Both beams were vertically polarized except for the circular dichroism measurement. The reflected probe beam was dispersed by a transmissive grating (300 gr/mm), focused by a lens (focal length = 5 cm), and then detected by a linear camera (Coptonix, S11639-01). The pump beam was chopped at 445 Hz, and the pump-probe signal $\Delta R/R$ was averaged for 3072 frames at each time delay point. Group delay dispersion induced by optical components is post-corrected during the analysis. The exact values of these parameters (pump energy, pump beam size, integration time, etc.) were varied for different measurements to optimize the signal. The pump-probe setup is shown in Figure S2. All measurements are done at 100 K unless otherwise noted.

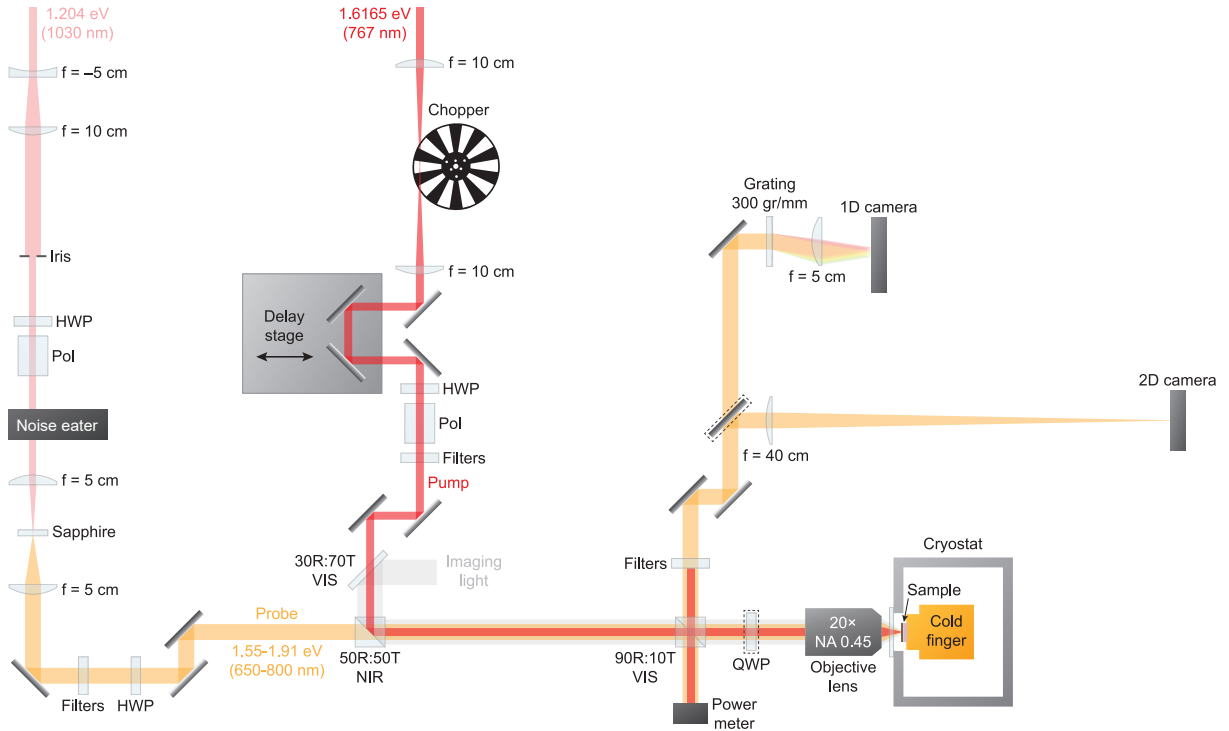

**Figure S2.** Collinear two-color pump-probe microscopy setup. Pump (1.6165 eV) and probe (1.55-1.91 eV) beams are collinear in the excitation path. A 750-nm long-pass filter is placed in the pump path, a combination of 650-nm short-pass and 800-nm long-pass filters is placed in the probe path, and a 750-nm short-pass filter is placed in the detection path. A quarter-wave plate (QWP) is placed only in circular dichroism measurement. A half-wave plate (HWP) and polarizer (Pol) are used to control the pump power.

We used a tightly focused probe beam (FWHM = 1.6 μm) and a large pump beam (FWHM = 6.2 μm) so that the beam wandering (~1 μm peak to peak) does not affect the effective pump fluence in the probed region. The large pump beam also allowed us to approximate its Gaussian profile $I(x,y) = I_0 e^{-(x^2+y^2)/2\sigma^2}$ to a flat-top profile with a corresponding area $2\pi\sigma^2$. We used the following conversion to calculate pump fluence: $F_{pump} = P_{pump}/2\pi\sigma^2 f_{rep}$, where $P_{pump}$ is the average pump power and $f_{rep}$ is the laser repetition rate. The number of incident pump photons is calculated as $n_{pump} = F_{pump}/E_{pump}$, where $E_{pump} = 1.616$ eV is the pump-photon energy. The pump and probe beam profiles are measured with a gray-scale CMOS camera (shown in Figure S3).

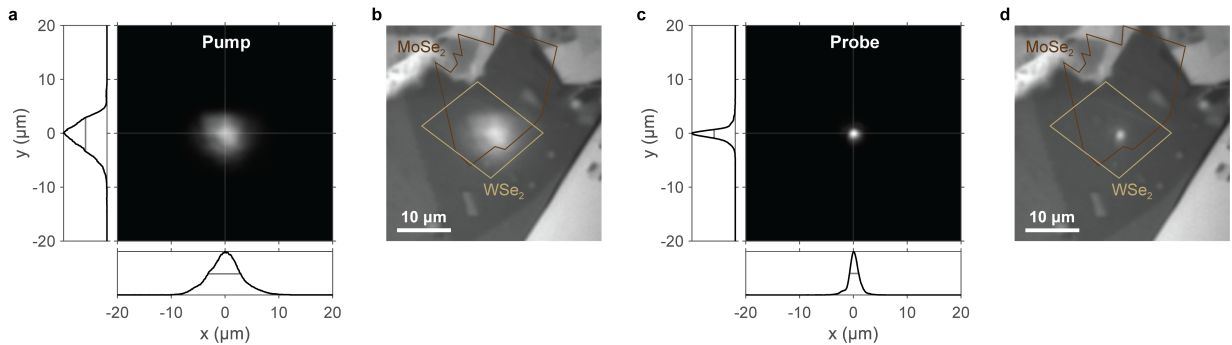

**Figure S3.** Pump and probe beam profiles. (a) An image of the pump beam on the sample and its linecuts at $x = 0$ and $y = 0$. (b) A white-light image with the pump beam. (c) An image of the probe beam on the sample and its linecuts at $x = 0$ and $y = 0$. (d) A white-light image with the probe beam.

## 3. Absorption calibration using transfer-matrix calculation

With measured layer dimensions and known dielectric constants, we simulated the reflection contrast spectrum of the $MoSe_2/hBN/WSe_2$ device using transfer-matrix calculations. Starting with the dimensions estimated from AFM measurements of FLG and hBN flakes, we varied the dielectric constants of $MoSe_2$ and $WSe_2$ layers with parameters such as the oscillator strength, peak energy, and linewidth, until the calculated reflection contrast spectrum matched the experimental one. From the fit (orange line in Figure 1e), the central energies and linewidths of A exciton peaks are determined as $E_{MoSe_2} = 1.6217$ eV, $\Gamma_{MoSe_2} = 12.0$ meV, $E_{WSe_2} = 1.6923$ eV, and $\Gamma_{WSe_2} = 13.8$ meV.

An absorption coefficient $\alpha$ is defined as the fraction of the power absorbed in a unit length of the medium. If the beam is propagating in the $z$ direction and the incident intensity is $I_0$, then the intensity at position $z$ is

$$I(z) = I_0 e^{-\alpha z}. \tag{S2}$$

Considering the complex refractive index $\tilde{n} = n + i\kappa$, the electric-field propagation is described as

$$\mathcal{E}(z,t) = \mathcal{E}_0 e^{i(kz-\omega t)} = \mathcal{E}_0 e^{i(\tilde{n}\omega z/c - \omega t)} = \mathcal{E}_0 e^{i(n\omega z/c - \omega t)} e^{-\kappa\omega z/c}. \tag{S3}$$

By comparing the exponential decay constants in Eqs. (S2) and (S3), considering that the optical intensity is $I \propto \mathcal{E}\mathcal{E}^*$, the absorption coefficient can be derived as $\alpha = 2\kappa\omega/c$. Therefore, the imaginary refractive index $\kappa$ is directly related to the absorption coefficient $\alpha$. By modeling complex dielectric constants $\tilde{\epsilon} = \epsilon_1 + i\epsilon_2$ with Lorentzian oscillators and fitting them to an experimental reflection contrast spectrum, we can estimate the value of imaginary refractive index

$$\kappa = \sqrt{\frac{-\epsilon_1 + \sqrt{\epsilon_1^2 + \epsilon_2^2}}{2}} \tag{S4}$$

and, subsequently, the absorption coefficient $\alpha = 2\kappa\omega/c$. The total amount of absorption by a monolayer with thickness $d$ at an angular frequency $\omega$ is then estimated as

$$A(\omega) = \frac{I_0 - I(d)}{I_0} = 1 - e^{-\alpha d} = 1 - e^{-2\kappa\omega d/c}. \tag{S5}$$

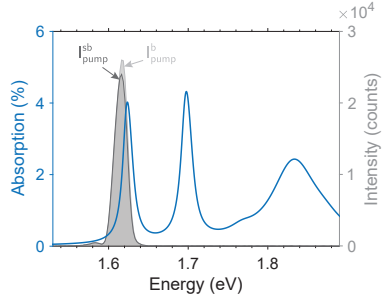

**Figure S4.** Estimation of pump absorption $A_{\text{tot}}$. Extracted sample absorption (from the reflection contrast spectrum fitting shown in Figure 1e) is plotted in blue (left axis). The pump spectra in the sample ($I_{\text{pump}}^{\text{sb}}$) and background ($I_{\text{pump}}^{\text{b}}$) regions are plotted in dark gray and light gray, respectively (right axis).

Alternatively, we can simply perform a transfer-matrix calculation to estimate the power reflectance $\mathcal{R}$ and power transmittance $\mathcal{T}$ of the sample ("sb") and background ("b") regions using the extracted complex dielectric constants (from the fit to the measured reflection contrast spectrum), and then calculate the absorption in each region as $A^{\text{sb}} = 1 - \mathcal{R}^{\text{sb}} - \mathcal{T}^{\text{sb}}$ and $A^{\text{b}} = 1 - \mathcal{R}^{\text{b}} - \mathcal{T}^{\text{b}}$. The monolayer absorption is then calculated as $A = A^{\text{sb}} - A^{\text{b}}$. We find that both methods yield very similar values of monolayer absorption $A(\omega)$.

Due to the sharp spectral features of both the pump and exciton peaks, the estimation of the total pump absorption has an additional complication (see Figure S4). To take the spectral mismatch into account, we calculated the total pump absorption $A_{\text{tot}}$ as

$$A_{\text{tot}} = \frac{\int I_{\text{pump}}(E) A(E) dE}{\int I_{\text{pump}}(E) dE} = 0.0196, \tag{S6}$$

where $I_{\text{pump}}(E)$ is the pump spectrum and $A(E)$ is the sample absorption spectrum extracted from the transfer-matrix calculation (both shown in Figure S4). Note that the value of $A_{\text{tot}}$ varies not only by the mismatch between the pump spectrum and sample resonance but also by the vertical stacking structure and its dielectric constants, which affect the strength of the local field at the TMD layer. The measured pump fluence $F_{\text{pump}}$ was converted to the incident photon density $n_{\text{pump}}$, and then converted to the pump-induced $\text{MoSe}_2$ exciton density $n_{\text{m}} = A_{\text{tot}} n_{\text{pump}}$. Even if we use a saturable absorber model

$$n_{\text{m}} = \frac{n_{\text{Mott}}}{n_{\text{pump}} + n_{\text{Mott}}/A_{\text{tot}}} \cdot n_{\text{pump}} \tag{S7}$$

using the excitonic Mott density $n_{\text{Mott}}$, the density-dependent trends shown in Figure 5 do not change significantly.

# 4. Transferred hole density calibration using gate-dependent reflection contrast

The transferred hole density $n_{\text{w}}$ in the $\text{WSe}_2$ layer can be estimated from the saturation of the exciton absorption peak at $E_x$ or from the appearance of the trion absorption peak at $E_t$ as a function of the applied back-gate voltage $V_{\text{bg}}$. We measured gate-dependent reflection contrast spectra at 4 K (Figure S5a), which show well-defined exciton and trion features on both the WSe$_2$-hole-doped side and the MoSe$_2$-electron-doped side. Vertical linecuts at various back-gate voltages are shown in Figure S5b, and horizontal linecuts at $E_x$ and $E_t$ are shown in Figure S5c. The WSe$_2$ hole density $n_{\mathrm{w}}$ (the $x$ axis of Figure S5c) is calibrated using a capacitor model with $\epsilon_{\mathrm{hBN}} = 4.2 \pm 0.4$ and $d_{\mathrm{hBN}} = 80$ nm (measured by an AFM). In the regime we are concerned with, the absorption change is linear with $n_{\mathrm{w}}$ (fit results to linear dependence are shown with block dotted lines in Figure S5c). This way, the saturation of the exciton peak (or the appearance of the trion peak) can be correlated with $n_{\mathrm{w}}$. The linear fit is valid for the results shown in Figures 2-5, where the transferred hole density is well below the Mott density (the change of WSe$_2$ exciton absorption induced by the pump is up to $|\Delta A/A|_{\mathrm{max}} \sim 12\%$). Because the intraband relaxation is ultrafast (<100 fs), we expect that the effect of gate-injected holes and the effect of transferred holes are not different at the ps time scale.

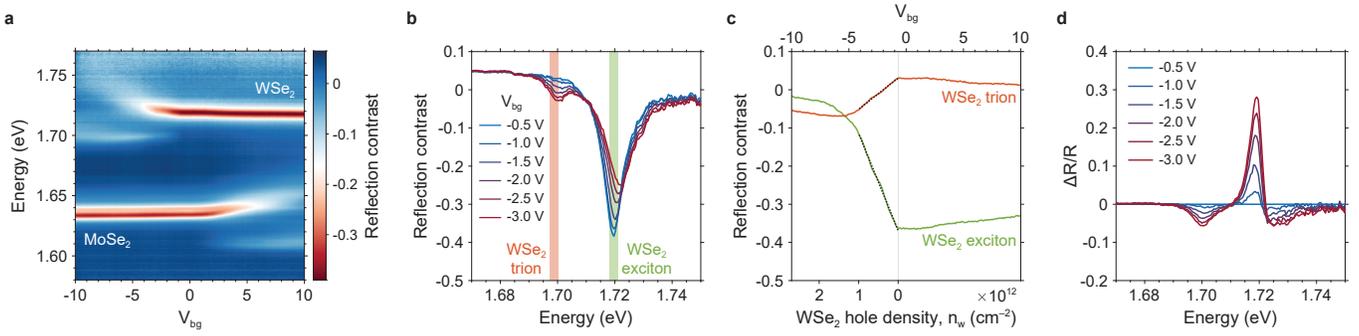

**Figure S5.** Estimation of transferred hole density in WSe$_2$. (a) Gate-dependent reflection contrast spectrum at 4 K. (b) Linecuts of (a) near the WSe$_2$ exciton energy at six different back-gate voltages ($V_{\mathrm{bg}}$). (c) Reflection contrast at the WSe$_2$ exciton energy (green, 1.7195 eV) and at the WSe$_2$ trion energy (orange, 1.6989 eV). Black dotted lines are fit results to a linear dependence. (d) Simulated pump-probe spectra using reflection contrast spectra at different gate voltages, which show a negative dip at the WSe$_2$ trion energy and a positive peak at the WSe$_2$ exciton energy.

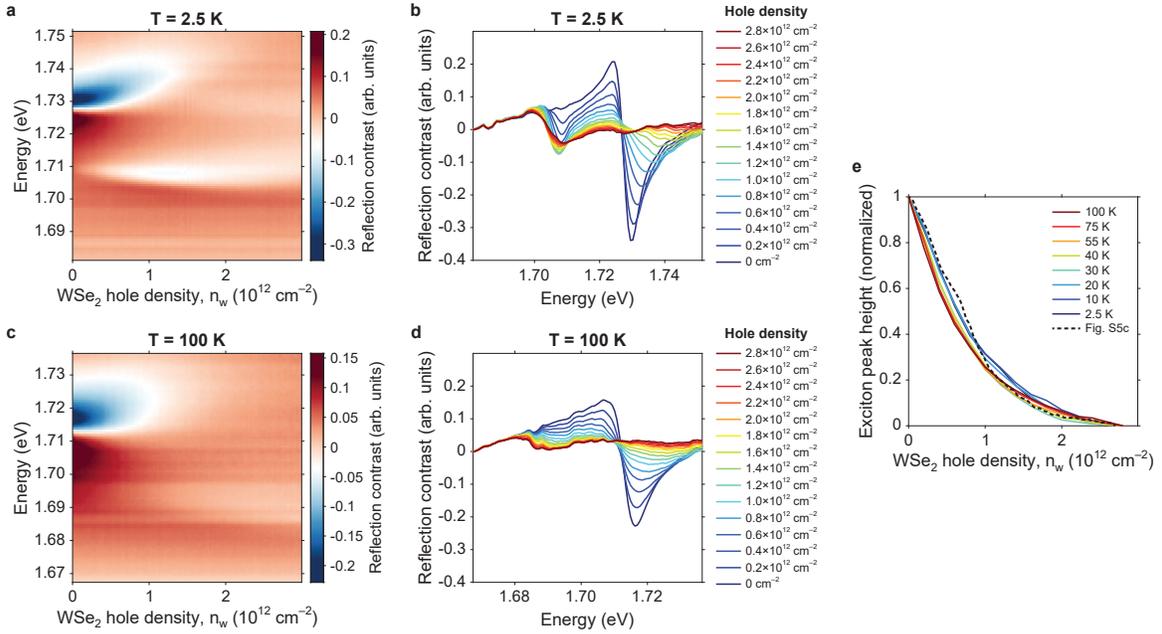

**Figure S6.** WSe$_2$ hole density ($n_{\mathrm{w}}$) calibration curves at different temperatures. (a) Gate-dependent reflection contrast spectra of MoSe$_2$/3 nm hBN/WSe$_2$ at 2.5 K. The MoSe$_2$ layer is kept intrinsic, and only the WSe$_2$ hole density is varied. (b) Linecuts of (a) at hole densities from 0 cm$^{-2}$ up to $2.8 \times 10^{12}$ cm$^{-2}$ at 2.5 K. (c) Gate-dependent absorption spectra at 100 K. (d) Linecuts of (c) at hole densities from 0 cm$^{-2}$ up to $2.8 \times 10^{12}$ cm$^{-2}$ at 100 K. (e) The normalized exciton peak height (1 = maximum exciton peak, 0 = minimum exciton peak) at temperatures from 2.5 K to 100 K. The normalized version of Figure S5c is also plotted with a black dashed line.

Using another gateable device, we confirmed that the calibration (exciton peak saturation vs. $n_{\mathrm{w}}$) is temperature independent from 2.5 K up to 100 K. Representative gate-dependent reflection contrast spectra and their linecuts are shown in Figures S6a-d, and the calibration curves at different temperatures are shown in Figure S6e. The normalized exciton peak height as a function of $n_w$ shows that the exciton peak saturates at the same rate for a wide range of temperatures. Although the estimation of uncertainties in $n_w$ is nontrivial, we used the hBN dielectric constant $\epsilon_{hBN} = 4.2 \pm 0.4$ and the uncertainties in the calibration curves to estimate that the overall uncertainty of $n_w$ does not exceed 20%.

We also simulated the measured pump-probe spectra (shown in Figures 3c,d) using gate-dependent reflection contrast spectra and plotted in Figure S5d. The pump-off spectrum ($R_{pump\ off}$) is simulated by the reflection contrast spectrum at the charge-neutral point ($-0.5$ V), and the pump-on spectrum ($R_{pump\ on}$) is simulated by the reflection contrast spectra at various gate voltages from $-1$ V to $-3$ V. The resulting pump-probe spectra, $\Delta R/R = (R_{pump\ on} - R_{pump\ off})/R_{pump\ off} \approx (R - R_{-0.5\ V})/R_{-0.5\ V}$, show the saturation of the exciton peak ($\Delta R/R > 0$ at $E_x$) and the appearance of the trion peak ($\Delta R/R < 0$ at $E_t$), similar to the measured pump-probe spectra shown in Figures 3c,d.

## 5. Results of coupled rate equations fitting

Here, we provide a few different fitting results of Eqs. (1) and (2) in the main text when different starting parameters and bounds are used. In particular, the goodness of fit and the fitted charge transfer time $\tau_t = 1/\gamma_t$ were dependent on the starting value of the MoSe$_2$ exciton lifetime $\tau_m = 1/\gamma_m$. However, the estimation of $\tau_m$ is difficult in the device used in the main text (MoSe$_2$/1 nm hBN/WSe$_2$) because the exciton decay process competes with hole transfer processes at the similar time scale. For this reason, we used another device that showed a negligible charge transfer process (MoSe$_2$/3 nm hBN/WSe$_2$) to estimate the value of $\tau_m$. Figure S7a shows a transient reflection spectrum measured at 100 K with the pump energy at 1.823 eV and a very low pump fluence. The carrier relaxation from the pump energy (1.823 eV) to the exciton states (1.704 eV and 1.622 eV) is known to be sub-100 fs (Ref. 28 of the main text), so the subsequent dynamics are dominated by the exciton population decay. The MoSe$_2$ exciton decay can be estimated by the linecut at 1.622 eV (Figure S7b), which is fitted to a double-exponential decay function $\Delta R/R = Ae^{-t/\tau_1} + Be^{-t/\tau_2} + C$ with $\tau_1 = 4.2$ ps, $\tau_2 = 119$ ps, and $A/B = 1.6$. On the other hand, the WSe$_2$ exciton decay can be estimated by the linecut at 1.704 eV (Figure S7c), which is fitted to the same function with $\tau_1 = 6.0$ ps, $\tau_2 = 133$ ps, and $A/B = 0.5$. The double-exponential functional form is typical when the thermal excitation is significant and leads to the exciton population outside the light cone. Although it is difficult to distinguish the radiative and nonradiative components from this pump-probe measurement, the effective MoSe$_2$ exciton lifetime $\tau_m$ can be well estimated to be ~120 ps.

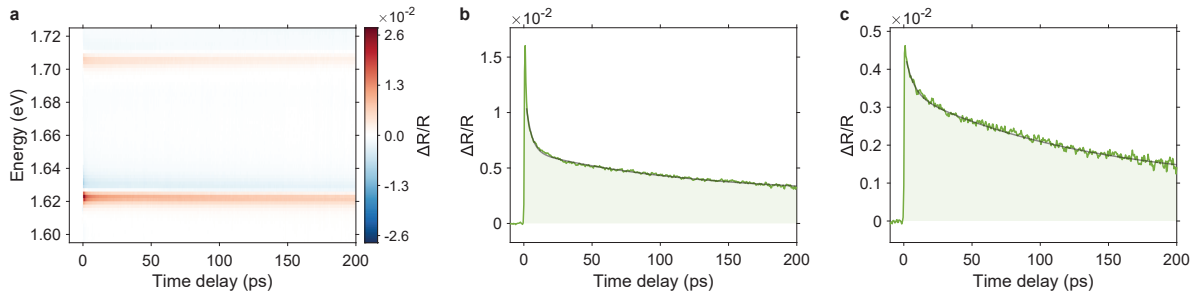

**Figure S7.** (a) Pump-probe spectrum of MoSe$_2$/3 nm hBN/WSe$_2$ measured with a pump at 1.823 eV. (b) A linecut at 1.622 eV showing the MoSe$_2$ exciton dynamics. (c) A linecut at 1.704 eV showing the WSe$_2$ exciton dynamics. Fits to double-exponential decay functions are shown with gray lines.

- For the results shown in Figures S8a-c, which are identical to that shown in Figure 5, the starting value of $\tau_m$ was set to 100 ps but was allowed to be varied up to 30% at different excitation densities. The density-dependent exciton lifetime accounts for nonlinear processes that are not fully captured by the exciton-exciton annihilation term $\frac{1}{2}\gamma_{mm}n_m^2$. Other fit parameters are $\gamma_w = (4\ ns)^{-1}$, $\gamma_{mm} = 0.27\ cm^2/s$, $\gamma_t = (500\ ps)^{-1}$, and $\alpha = 2.3\%$.

- For the results shown in Figures S8d-f, the starting value of $\tau_m$ was allowed to be varied from 200 ps to 10 ps at different excitation densities. This gave the best fitting across all densities—both the slow rise at the low excitation densities and the fast rise at the high excitation densities are well captured. Other fit parameters are $\gamma_w = (5\ ns)^{-1}$, $\gamma_{mm} = 0.23\ cm^2/s$, $\gamma_t = (500\ ps)^{-1}$, and $\alpha = 2.9\%$. However, such a wide range of $\tau_m$ values may not be physical.

- For the results shown in Figures S8g-i, the final value of $\tau_m$ was fixed for all excitation densities. When the starting value of $\tau_m$ was 100 ps, the final value of $\tau_m = 61$ ps gave the best fit to all curves. The resulting parameters are $\gamma_w = (4\text{ ns})^{-1}$, $\gamma_{mm} = 0.47\text{ cm}^2/\text{s}$, $\gamma_t = (190\text{ ps})^{-1}$, and $\alpha = 1.6\%$. However, the fits at both the low and high excitation densities are poor, not being able to reproduce the slow rise at the low excitation densities and fast rise at the high excitation densities.

- For the results shown in Figures S8j-l, the final value of $\tau_m$ was fixed for all excitation densities. When the starting value of $\tau_m$ was 200 ps, the final value of 175 ps gave the best fit to all curves. The resulting parameters are $\gamma_w = (3\text{ ns})^{-1}$, $\gamma_{mm} = 0.34\text{ cm}^2/\text{s}$, $\gamma_t = (500\text{ ps})^{-1}$, and $\alpha = 2.3\%$. The fits are better than Figures S8g-i at the low excitation densities but not as good as Figures S8a-f.

For different starting parameters, we consistently observed that the charge transfer rate is $\gamma_t = (500\text{ ps})^{-1}$. Our estimate is consistent with the measured transfer efficiency in the low-density limit, $\eta = \tau_m/(\tau_t + \tau_m)$, and the measured MoSe$_2$ exciton lifetime $\tau_m$ at 100 K (shown in Figure S7b).

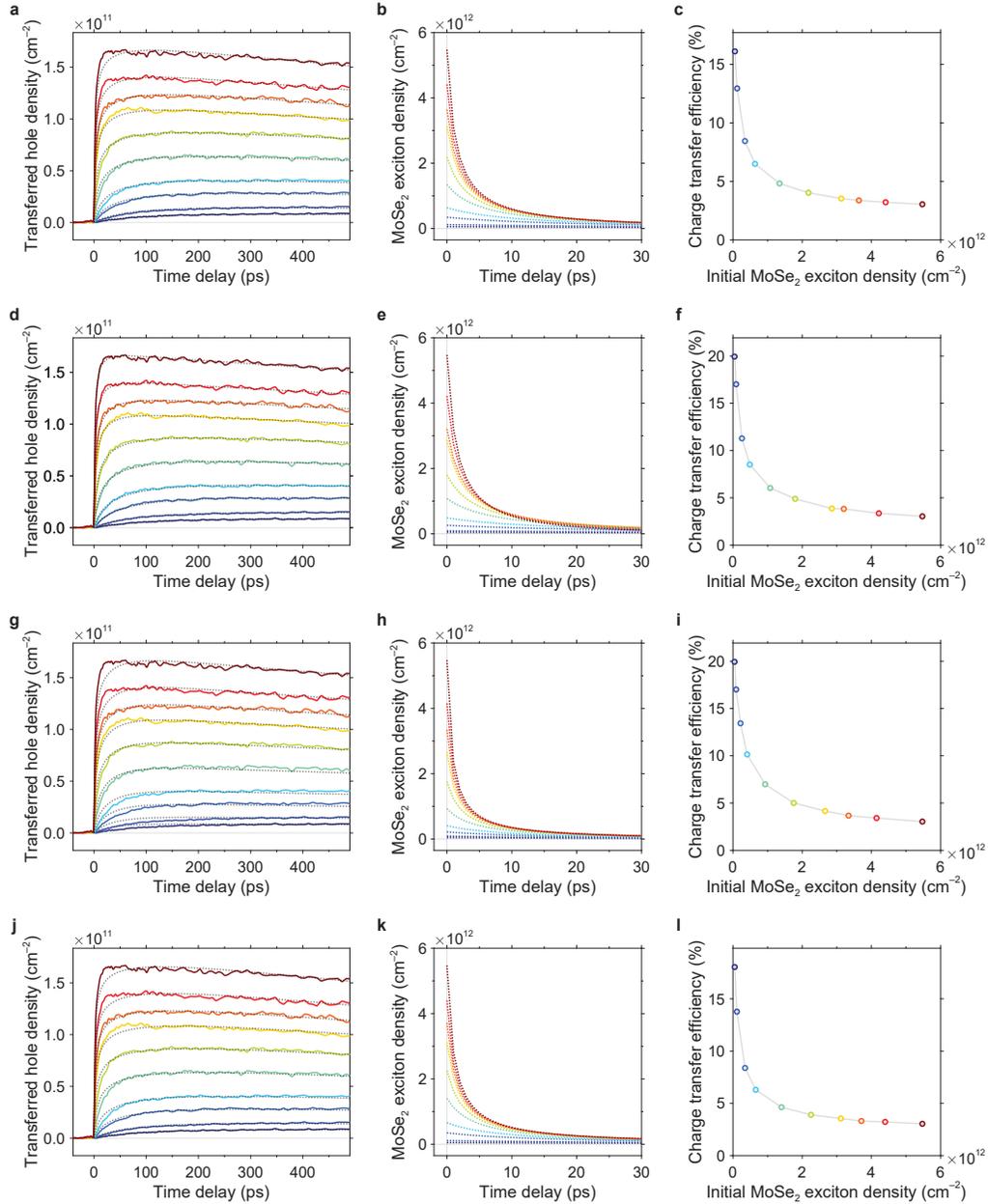

**Figure S8.** Results of the coupled rate equations fitting with different conditions.

## 6. Momentum mismatch between MoSe₂ and WSe₂ valleys

Interlayer excitons are not only separated in real space but also in momentum space depending on the twist angle and lattice mismatch between the two layers, which further modifies recombination pathways and dynamics. Using the lattice constants of MoSe₂ and WSe₂ layers, $a_m = 3.29$ Å and $a_w = 3.28$ Å, their momenta at K valleys from the $\Gamma$ point can be calculated as $k_{m,K} = 4\pi/3a_m = 1.273$ Å$^{-1}$ and $k_{w,K} = 4\pi/3a_w = 1.277$ Å$^{-1}$. The distance between the two points in the momentum space is given by

$$\Delta k = \sqrt{k_{m,K}^2 + k_{w,K}^2 - 2k_{m,K}k_{w,K}\cos\theta_{twist}}, \tag{S8}$$

where $\theta_{twist} = 19.5°$ for the device used in the main text. The momentum mismatch between the K valleys of MoSe₂ and WSe₂ is $\Delta k = 0.43$ Å$^{-1}$, while the momentum mismatch between the MoSe₂ K valley and WSe₂ K′ valley is 0.88 Å$^{-1}$. We performed a circular dichroism measurement where the pump beam was right-circularly polarized ($\sigma^+$) and the pump-probe spectra with right- and left-circular ($\sigma^+$ and $\sigma^-$) probe beams are compared, which shows that the transfer timescale and efficiency from K$_{MoSe_2}$ to K$_{WSe_2}$ valleys are similar to that from K$_{MoSe_2}$ to K′$_{WSe_2}$ valleys (within 10%).

The measurement of twist-angle-dependent charge transfer dynamics can potentially identify which momentum pathway it takes (either direct transfer from the MoSe₂ K valley to the WSe₂ K valley or via the Q valley that has a mixed character), while the temperature dependence can provide information about the population of phonons that connect the momentum and energy mismatch. Previous twist-angle-dependent or temperature-dependent measurements were not able to find any significant variations greater than the instrumental time resolution (10-100 fs). Pushing the instrument response limit would allow us to accurately compare the transfer timescales under different conditions. Using extremely short pulses (<10 fs), however, is not only technically challenging but also not ideal in the sense that the material response cannot keep up with the extremely short perturbation (in other words, the system does not interact strongly with the nonresonant part of the ultrashort broadband pump). Therefore, the slow and experimentally resolvable charge transfer in TMD/hBN/TMD offers new possibilities to investigate the charge transfer mechanism.